\documentclass[bibyear]{aa} 
\usepackage{epsfig}
\usepackage{amsmath}
\usepackage{subfigure}
\usepackage{natbib}
\usepackage{multicol}
\usepackage{multirow}
\usepackage{color}
\usepackage{float}
\usepackage{longtable}
\usepackage{graphicx}
\usepackage{epstopdf}
\usepackage{booktabs}
\usepackage{txfonts}

\newcommand{\jms}{J.~Mol.~Spectrosc.}   
\newcommand{\jmst}{J.~Mol.~Struct.}   

\newcommand{\kms}{km s$^{-1}$}

\newcommand{\prev}{Phys. Rev.}

\begin{document}

\title{Discovery of the C$_7$N$^-$ anion in TMC-1 and IRC\,+10216\thanks{Based on observations carried out
with the Yebes 40m telescope (projects 19A003,
20A014, 20D023, and 21A011). The 40m
radiotelescope at Yebes Observatory is operated by the Spanish Geographic 
Institute
(IGN, Ministerio de Transportes, Movilidad y Agenda Urbana).}}

\author{
J.~Cernicharo\inst{1},
J.~R.~Pardo\inst{1},
C.~Cabezas\inst{1},
M.~Ag\'undez\inst{1},
B.~Tercero\inst{2,3},
N.~Marcelino\inst{2},
R.~Fuentetaja\inst{1},
M.~Gu\'elin\inst{4}, and
P.~de~Vicente\inst{3}
}

\institute{Grupo de Astrof\'isica Molecular, Instituto de F\'isica Fundamental (IFF-CSIC),
C/ Serrano 121, 28006 Madrid, Spain\\ \email jose.cernicharo@csic.es 
\and Centro de Desarrollos Tecnol\'ogicos, Observatorio de Yebes (IGN), 19141 Yebes, Guadalajara, Spain
\and Observatorio Astron\'omico Nacional (OAN, IGN), Madrid, Spain
\and Institut de Radioastronomie Millim\'etrique, 300 rue de la Piscine, F-38406,
  Saint Martin d'H\`eres, France
}

\date{Submitted 29/12/2022; Accepted 30/1/2023}

\abstract{
We report on the discovery of the C$_7$N$^-$ anion towards the starless core
TMC-1 and towards the carbon-rich evolved star IRC\,+10216. 
We used the data of the 
QUIJOTE$^1$
line survey towards TMC-1
and found six lines in perfect harmonic frequency relation
from $J$=27-26 up to $J$=32-31. The frequency of the lines can be reproduced
with a rotational constant and a distortion constant of $B$=582.68490$\pm$0.00024 MHz
and $D$=4.01$\pm$0.13 Hz, respectively. The standard deviation
of the fit is 4 kHz. 
Towards IRC\,+10216, we identify 17 lines from $J$=27-26 up to $J$=43-42;  their frequencies are also in harmonic relation, providing
$B$=582.6827$\pm$0.00085 MHz 
and $D$=3.31$\pm$0.31 Hz. The nearly exact coincidence of the rotational and
distortion constants in both sources points unambiguously to a common molecular carrier. Taking
into account the chemical peculiarities of both sources, the carrier could be a radical or 
an anion. The radical can be discarded, as the observed lines belong to a singlet species. Hence, 
the most
plausible carrier is an anion. High-level ab initio calculations indicate that
C$_7$N$^-$, for which we compute a rotational constant of $B$=582.0 MHz and a dipole moment
of 7.5\,D, 
is the carrier of the lines in both sources. We predict the neutral C$_7$N to have
a ground electronic state $^2\Pi$ and a dipole moment of $\sim$1\,D. Because of this low value
of $\mu$ and to its much larger rotational partition function, its lines are expected to
be well below the sensitivity of our data for both sources.
}

\titlerunning{C$_7$N$^-$ in TMC-1 and IRC\,+10216}
\authorrunning{Cernicharo et al.}

\maketitle

\section{Introduction}

The presence
of carbon-chain negative ions in space was predicted on the grounds that electron
radiative attachment is efficient for open-shell molecules with large electron
affinities \citep{Dalgarno1973,Sarre1980,Herbst1981}. The predicted abundance ratio
between an anion and its corresponding neutral species increases with the size of
the molecule, which is a consequence of the increasing density of vibrational states
as the number of atoms in the molecule increases.

The first anion detected in space, C$_6$H$^-$, was observed towards TMC-1 \citep{McCarthy2006}.
Lines from this species were already reported as unidentified
features in the line survey of IRC\,+10216 performed with the Nobeyama 45m telescope
by \citet{Kawaguchi1995}. 
\citet{Aoki2000}
suggested that the carrier of these lines was C$_6$H$^-$ from \emph{ab initio} calculations, which
was finally confirmed by
the laboratory observations of \citet{McCarthy2006}. 
The presence of this anion in space attracted significant attention and motivated searches
for other hydrocarbon anions in interstellar and circumstellar clouds. C$_4$H$^-$
was first discovered in the circumstellar cloud IRC\,+10216 by \citet{Cernicharo2007}
and then in the interstellar clouds L1527, Lupus-1A, and TMC-1 
\citep{Agundez2008,Sakai2008,Sakai2010,Cordiner2013}. 
Following the observation of C$_8$H$^-$ in the laboratory \citep{Gupta2007},
this anion was found in TMC-1 \citep{Brunken2007}, IRC\,+10216 \citep{Kawaguchi2007, Remijan2007}, and Lupus-1A \citep{Sakai2010}.
\citet{Remijan2023} recently announced the detection of C$_{10}$H$^-$ in TMC-1. Although we searched for 
it and other anions, we cannot confirm any line of this species in the 
QUIJOTE\footnote{\textbf{Q}-band \textbf{U}ltrasensitive \textbf{I}nspection \textbf{J}ourney to the \textbf{O}bscure \textbf{T}MC-1 \textbf{E}nvironment} 
line survey, probably due to the 
high-$J$ levels involved in the rotational transitions of C$_{10}$H$^-$
in our data ($J_u\ge$52 and $E_u$> 40\,K).

\begin{figure*}
\centering
\includegraphics[width=0.88\textwidth,angle=0]{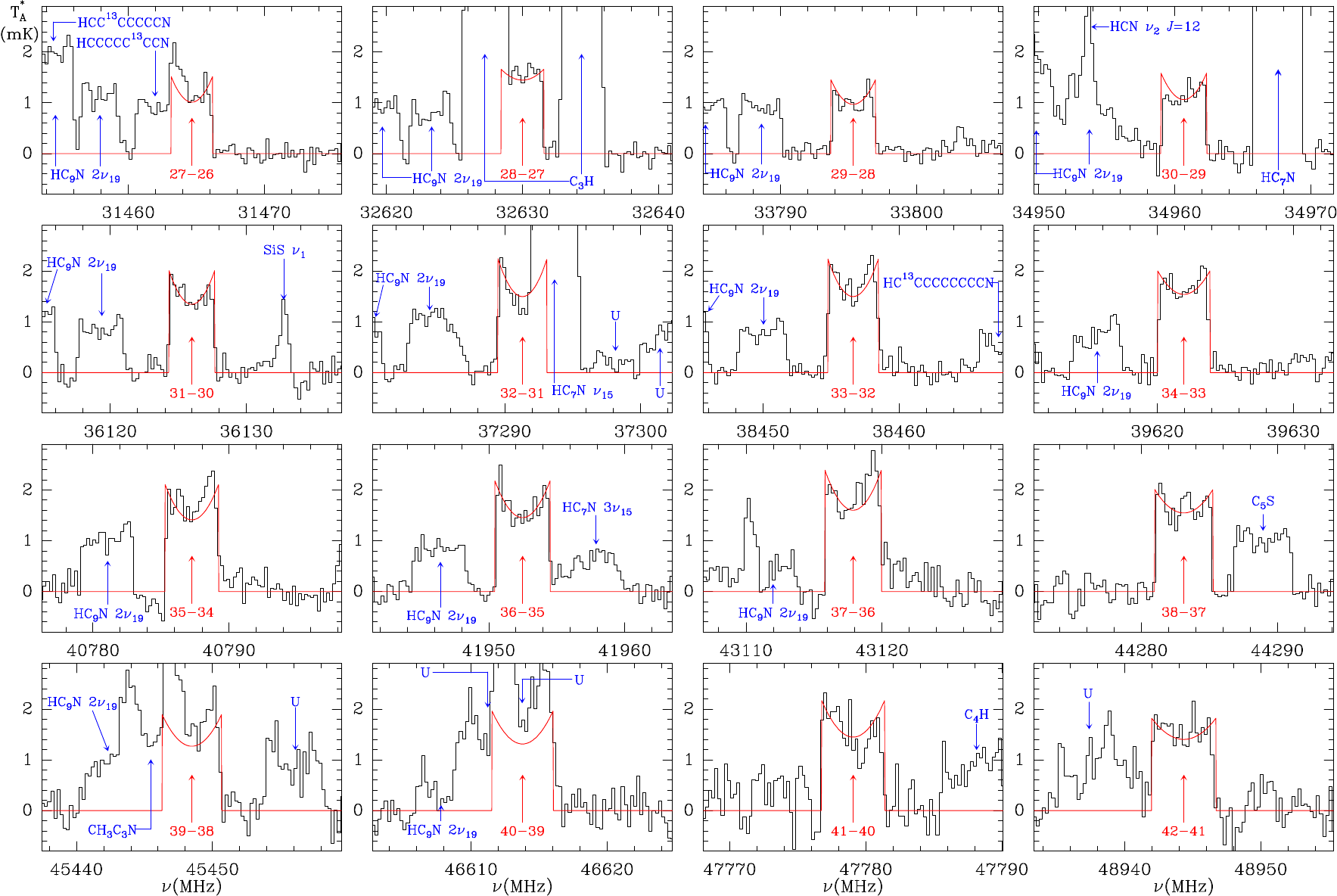}
\caption{Observed lines of C$_7$N$^-$ towards IRC\,+10216. 
Line parameters are given in Table \ref{line_parameters_irc}.
The abscissa corresponds to the rest frequency assuming a local standard of rest velocity of $-$26.5
km s$^{-1}$. 
The ordinate is the antenna temperature corrected for atmospheric and telescope losses in milliKelvin.
The red lines show the fitted line profiles.
}
\label{fig_irc}
\end{figure*}

\begin{figure}
\centering
\includegraphics[width=0.43\textwidth]{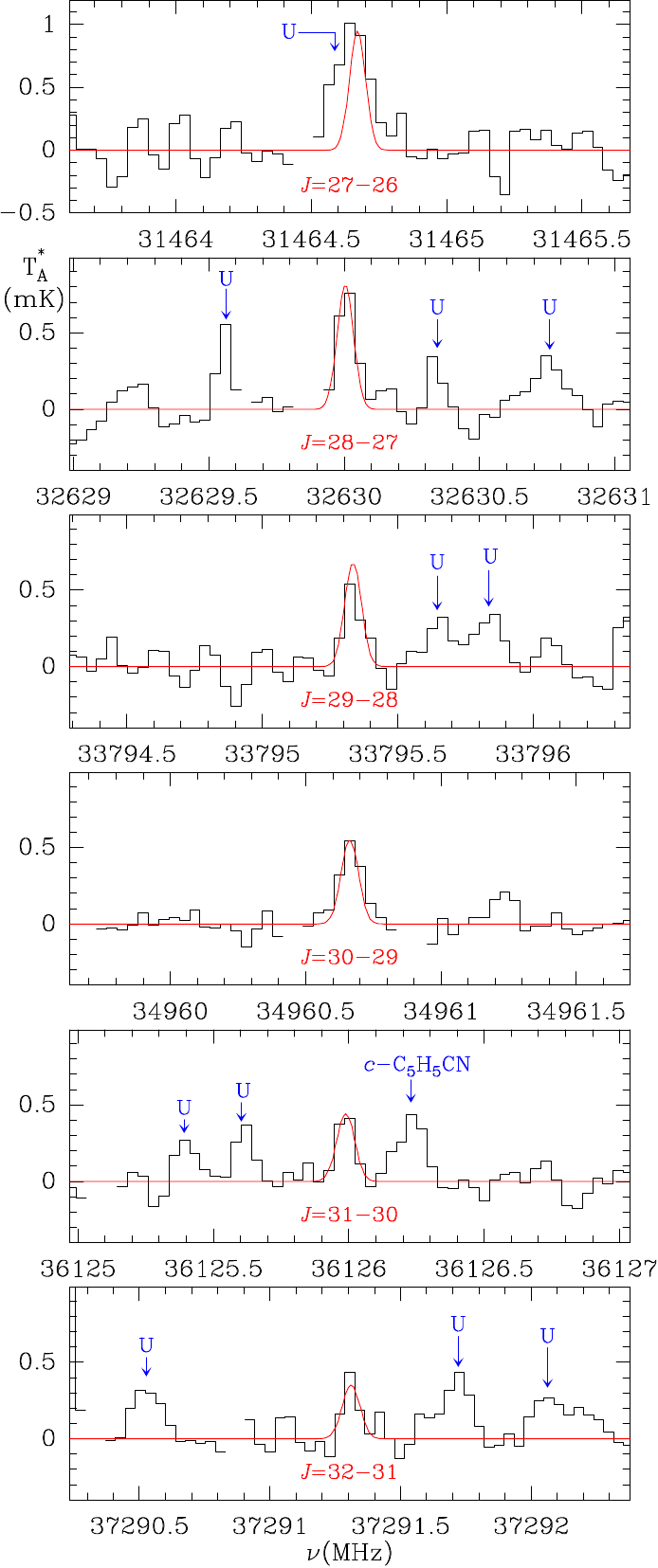}
\caption{Observed lines of C$_7$N$^-$ towards TMC-1. 
Line parameters are given in Table \ref{line_parameters_tmc1}.
The abscissa corresponds to the rest frequency assuming a local standard of rest velocity of $+$5.83
km s$^{-1}$. 
The ordinate is the antenna temperature corrected for atmospheric and telescope losses in milliKelvin.
The red line shows the synthetic spectrum derived for
T$_{rot}$=6\,K and N(C$_7$N$^-$)=5$\times$10$^{10}$ cm$^{-2}$.
Blank channels correspond to negative features produced
in the folding of the frequency-switching data.
}
\label{fig_tmc1}
\end{figure}

The nitrile anions CN$^-$, C$_3$N$^-$, and C$_5$N$^-$ were first detected in the circumstellar envelope of the carbon-rich 
star IRC\,+10216 \citep{Agundez2010,Thaddeus2008,Cernicharo2008,Cernicharo2020}. 
CN$^-$ and C$_3$N$^-$ were detected thanks to accurate laboratory
frequencies \citep{Gottlieb2007,Thaddeus2008,Amano2008}. However, the assignment of C$_5$N$^-$
was based on \emph{ab initio} calculations by \citet{Botschwina2008} and \citet{Aoki2000}. Although C$_5$N$^-$ was the best candidate, 
the lack of precise laboratory frequencies prevented metal-bearing molecules being ruled out, such as MgC$_3$N and MgC$_4$H. Later on, 
these two molecules were identified in IRC\,+10216 based on \emph{ab initio} calculations \citep{Cernicharo2019}. Moreover, a series of lines with the same rotational constant as that assigned to C$_5$N$^-$ in IRC\,+10216 were recently observed in TMC-1 \citep{Cernicharo2020}, which ruled out metal-bearing molecules and provided strong support in favour of C$_5$N$^-$.

There is controversy over the formation of C$_n$N$^-$ anions through radiative 
electron attachment to C$_n$N radicals because calculated rate coefficients can differ by orders of 
magnitude depending on the study \citep{Petrie1997,Walsh2009,Khamesian2016,Millar2017}. Hence, the detection of nitrile anions 
and the determination of their abundances in different astronomical environments
are an important step forward in understanding the chemistry of anions in space.

In this letter, we present the discovery of C$_7$N$^-$ in the cold dark core TMC-1 and in the external
layers of the circumstellar envelope of the carbon-rich star IRC\,+10216. 
The assignment is based on the excellent agreement between the derived
rotational constant and that from high-level \emph {ab initio} calculations performed in this work (see also \citealt{Botschwina2008}). The assignment to C$_7$N$^-$ is strengthened by the lack of other plausible candidates common to both sources.

\section{Observations} \label{observations}
New receivers built within the Nanocosmos\footnote{ERC grant ERC-2013-Syg-610256-NANOCOSMOS.\\
https://nanocosmos.iff.csic.es/} project
and installed at the Yebes 40m radiotelescope were used
for observations of TMC-1
($\alpha_{J2000}=4^{\rm h} 41^{\rm  m} 41.9^{\rm s}$ and $\delta_{J2000}=
+25^\circ 41' 27.0''$) and IRC\,+10216
($\alpha_{J2000}=9^{\rm h} 47^{\rm  m} 57.36^{\rm s}$ and $\delta_{J2000}=
+13^\circ 16' 44.4''$). 
The observations of TMC-1 belong to the QUIJOTE$^1$ line survey \citep{Cernicharo2021a}
and those of IRC\,+10216 to the Nanocosmos$^2$
survey of evolved stars \citep{Pardo2022}.
A detailed description of the telescope, receivers, and backends is 
given by \citet{Tercero2021}. Briefly, the receiver consists of two cold, high-electron-mobility transistor amplifiers covering the
31.0-50.3 GHz band with horizontal and vertical             
polarisations. Receiver temperatures in the runs achieved during 2020 vary from 22 K at 32 GHz
to 42 K at 50 GHz. Some power adaptation in the down-conversion chains reduced
the receiver temperatures during 2021 to 16\,K at 32 GHz and 25\,K at 50 GHz.
The backends are $2\times8\times2.5$ GHz fast Fourier transform spectrometers
with a spectral resolution of 38.15 kHz,
providing coverage of the whole Q-band in both polarisations. For the observations of TMC-1,
the original spectral resolution was used, while for IRC\,+10216, the data have been smoothed
to 220 kHz, which corresponds to a velocity resolution 
of 1.7\,km\,s$^{-1}$ at 40 GHz. This resolution
is large enough to resolve the U-shaped lines of IRC\,+10216, which exhibit a full velocity width of
29 km\,s$^{-1}$ \citep{Cernicharo2000}.

All observations of TMC-1 were performed in the frequency 
switching mode with frequency throws of 8 and 10 MHz. Details 
of the survey and the data-analysis procedure are given in \citet{Cernicharo2021a,Cernicharo2022}. 
The observations of IRC\,+10216
were performed in position switching and are described by \citet{Pardo2022}. 
The main beam efficiency of the telescope varies from 0.6 at
32 GHz to 0.43 at 50 GHz. The intensity scale used in this work, namely antenna temperature
($T_A^*$), was calibrated using two absorbers at different temperatures and the
atmospheric transmission model ATM \citep{Cernicharo1985, Pardo2001}.
Calibration uncertainties of 10~\% are adopted. However, some systematic effects
were observed between the data acquired during 2022 and 2021. The intensity of the
lines in the earliest observations of QUIJOTE are typically 10\%-15\% weaker than those of
the most recent observations. No frequency effects are observed in the derived calibration
factors. This systematic change is probably related to the surface adjustement after
holography measurements at the end of 2021. The whole set of data has been recalibrated and co-added to
produce the QUIJOTE line survey used for this work. The new line intensities, and 
hence the column densities, are $\sim$15\% larger than those previously reported. The data for TMC-1 and IRC\,+10216
presented here correspond to 758 h and 696 h  of on-source
telescope time, respectively. All data were analysed using the GILDAS package\footnote{\texttt{http://www.iram.fr/IRAMFR/GILDAS}}.

\begin{table*}
\caption{Rotational and distortion constants for C$_7$N$^-$}
\label{rota}
\centering
\begin{tabular}{lcccc}
                 &    TMC-1$^a$       &  IRC\,+10216$^b$    & All data$^c$ & Ab initio$^d$\\
\hline
\hline
$B$ (MHz)   &  582.68492$\pm$0.00024& 582.68274$\pm$0.00085 &  582.68490$\pm$0.00022& 582.75 \\
$D$ (Hz)  &    4.01$\pm$0.13       &   3.31$\pm$0.31        &    4.00$\pm$0.12     &   4.3\\
\hline
N$_{lines}$      & 6            &   17         &   17        & \\
$J_{max}$        & 32           &   43         &   43        & \\
$\sigma$ (kHz)   & 4.0          &   61         &   43        & \\
\end{tabular}
\tablefoot{\\
\tablefoottext{a}{Parameters derived using the frequencies measured towards TMC-1 (see
                  Table \ref{line_parameters_tmc1}.} 
\tablefoottext{b}{Parameters derived using the frequencies measured towards IRC\,+10216 (see 
                  Table \ref{line_parameters_irc}).}
\tablefoottext{c}{Parameters derived using the frequencies of the lines $J$=27-26 up to $J$=32-31
measured in TMC-1, and $J$=33-32 up to $J$=43-42 measured in IRC\,+10216.}
\tablefoottext{d}{Computed rotational constant from ab initio calculations 
as described in the text (see also \citealt{Botschwina2008}).} 
}
\end{table*}

\section{Detection of C$_7$N$^-$} \label{results}
Line identification in this work was done using the catalogues 
MADEX \citep{Cernicharo2012}, CDMS \citep{Muller2005}, and JPL \citep{Pickett1998}. 
By November 2022, the MADEX code contained 6446 spectral
entries corresponding to the ground and vibrationally excited states ---together
with the corresponding isotopologues--- of 1746 molecules. 

The new data of IRC\,+10216 have
an unprecedented sensitivity of 0.1 mK in the $T_A^*$ scale and 
for a spectral resolution of 220 kHz, and the survey shows a forest of unidentified lines.
A series of doublets and triplets are found in very good harmonic relation at slightly 
higher frequencies than the rotational lines of HC$_9$N; their rotational constants are
around 291 MHz. These results will be reported in a forthcoming paper
(Pardo et al., in preparation). These lines correspond  to the $\nu_{19}$=2,3 vibrational 
levels of HC$_9$N. Their
quantum numbers go from $J_u$=54 up to $J_u$=80. Some of them are shown in Fig. \ref{fig_irc}. In addition, we
find that  an
extra line was detected every two transitions of the HC$_9$N $\nu_{19}$=2 transitions following
a perfect harmonic relation from $J$=27-26 up to $J$=42-43 with a rotational constant of 
$B$=582.6 MHz. The
possibility that these lines also belong to a vibrational excited state of HC$_9$N is immediately ruled
out because if $B$ were 291.3, then all lines with odd $J_u$  would be missing. The lines are
shown in Fig. \ref{fig_irc} and appear as single features; 
the carrier is therefore a $^1\Sigma$ molecule, or a $^2\Sigma$ molecule with a very small value of
$\gamma$. A fit to the observed lines provides the rotational and
distortion constants given in Table \ref{rota}. Taking into account the observed line profile of the
unblended lines, the
carrier has to be produced in the external layers of the circumstellar envelope of IRC\,+10216, that is, in
the region where chemistry  is controlled by the Galactic UV photons and where radicals, cyanopolyynes, 
carbon chains, metal-bearing species, and anions are detected (see e.g. Cernicharo et al. 2000).

In order to obtain additional insight into the carrier of the lines we checked if they are present in
the QUIJOTE$^1$ line survey of TMC-1. Six of them are clearly detected, corresponding to the transitions $J$=27-26
up to $J$=32-31. These are shown in Fig. \ref{fig_tmc1}. None of them show signs of
fine or hyperfine structure and they therefore definitively belong to a $^1\Sigma$ molecule. 
The rotational and distortion constant are given in Table \ref{rota} and agree very well with
those derived from the 17 lines of IRC\,+10216. Moreover, the common measured frequencies
in both sources are identical within the measured uncertainties for the IRC\,+10216 lines.
The carrier is therefore common to both sources and consequently we can exclude a
metal-bearing species or a vibrationally excited state as these are not expected in TMC-1. 
As we find no evidence of fine structure, we can also discard a radical as the carrier. 
Moreover, we can also discard an isotopologue of an 
abundant species such as HC$_7$N for which $B$=564 MHz, and C$_8$H for which $B$ is 587.264 MHz \citep{McCarthy1999}. An
isotopologue of C$_8$H can be discarded due to the fact that the carrier has to be a singlet state. The
rotational constants of the isotopologues of HC$_7$N have been measured in the laboratory 
\citep{McCarthy2000} and all them have rotational constants below that of the carrier.
Hence, we have to conclude that we
are dealing with a new unknown molecular species. Nevertheless, its structure has to be similar to that
of HC$_7$N and C$_8$H. An isomer of HC$_7$N can be excluded as HC$_6$NC, 
the most likely candidate, has 
been observed in the laboratory to have a rotational constant of 582.5203 MHz 
\citep{Botschwina1998} and is not detected in either TMC-1 or IRC+10216.
The other isomers, HC$_x$NC$_{7-x}$, are expected to have larger rotational constants, except
HNC$_7$ for which the expected rotational constant is around 567 MHz
\citep{Silva1997,Lee2000}. Therefore, the most plausible candidates are C$_7$N and C$_7$N$^-$. 
The former can be discarded
as its ground electronic state is predicted to be $^2\Pi$ with a $^2\Sigma$ state 
close in energy \citep{Botschwina1997}.
This leaves the anion C$_7$N$^-$ as the most likely carrier of the lines
in TMC-1 and IRC\,+10216. Lines of the previous member of the series, C$_5$N$^-$, have been found
in both sources \citep{Cernicharo2008,Cernicharo2020}. Ab initio calculations by \citet{Botschwina2008}
indicate a rotational constant for this anion of 582 MHz, very close indeed to the observed value,
and a dipole moment of 7.545 D.

We performed geometry ab 
initio calculations \citep{Werner2020} using the coupled cluster 
method with single, double, and perturbative triple excitations with an explicitly correlated 
approximation (CCSD(T)-F12; \citealt{Knizia2009}) and all electrons (valence and core) 
correlated together with the Dunning's correlation consistent basis sets with polarized 
core-valence correlation triple-$\zeta$ for explicitly correlated calculations 
(cc-pCVTZ; \citealt{Hill2010}). The values for the centrifugal distortion constant 
were calculated \citep{Frisch2016} using the MP2 perturbation theory method \citep{Moller1934} 
and the correlation is consistent with polarized valence triple-$\zeta$ basis set 
(cc-pVTZ; \citealt{Woon1993}). The values of $B$ and $D$ calculated for C$_7$N$^-$ 
were scaled by the corresponding experimental and calculated ratios obtained for HC$_7$N 
using the same level of theory.  For HC$_7$N, we obtained $B$=562.9 MHz and $D$=3.50 Hz, 
while for C$_7$N$^-$ the uncorrected values are $B$=581.6 MHz and $D$=3.70 Hz. The scaled 
values for C$_7$N$^-$ are presented in Table \ref{rota} and perfectly
match those measured
for the lines in TMC-1 and IRC\,+10216.
 
We therefore conclude that we have discovered the C$_7$N$^-$ anion in the interstellar and
circumstellar media. A fit to the lines observed in both sources provides
$B$=582.68495$\pm$0.0024 MHz and $D$=4.02$\pm$0.13 Hz (see Table \ref{rota}), which are
the recommended rotational parameters with which to predict the spectrum of C$_7$N$^-$.

\section{Discussion}
Assuming a source emission diameter of 80$''$ for C$_7$N$^-$ in TMC-1 \citep{Fosse2001}, 
a fit to the observed line profiles results in a rotational temperature of 6.0$\pm$0.5\,K
and a column density of (5.0$\pm$0.5)$\times$10$^{10}$ cm$^{-2}$. Adopting
an averaged column density of H$_2$ over the source of 10$^{22}$ cm$^{-2}$
\citep{Cernicharo1987}, the abundance of C$_7$N$^-$ is 5$\times$10$^{-12}$. 
The column density of C$_5$N$^-$ in TMC-1 is (2.6$\pm$0.9)$\times$10$^{11}$ cm$^{-2}$
\citep{Cernicharo2020}, and therefore the abundance ratio 
C$_5$N$^-$/C$_7$N$^-$ is $\sim$5. However, we note that due to the high energy
of the rotational levels associated to the observed transitions, small 
variations of the rotational temperature can be compensated by a significant
variation of the column density with a negligible variation of $\chi^2$. This means that,
in light of the lack of information regarding the intensity of low-$J$ transitions, T$_{rot}$
and N are strongly correlated.

For IRC\,+10216, we assume that the emitting region 
has a radius of 15$''$, which
corresponds to the observed spatial distribution of cyanopolyynes and radicals with ALMA
\citep{Agundez2017}. A rotation diagram of the observed intensities towards 
IRC\,+10216
provides a rotational temperature of 26.6$\pm$1.8\,K and a column density of 
(2.4$\pm$0.2)$\times$10$^{12}$
cm$^{-2}$ for C$_7$N$^-$. The column density of C$_5$N$^-$ was previously 
derived in this source from lines
observed at 3\,mm to be 3.4$\times$10$^{12}$ cm$^{-2}$ \citep{Cernicharo2008}. 
However,
several lines of C$_5$N$^-$ have been observed in the Q-band in our survey 
with the Yebes\,40m
telescope and those at 3\,mm observed with the IRAM\,30m telescope have been
significantly improved \citep{Cernicharo2020}. From all these data, we 
derived a column density of C$_5$N$^-$
in IRC\,+10216 from a rotation diagram analysis, which provides 
N(C$_5$N$^-$)=
(5.7$\pm$0.3)$\times$10$^{12}$ cm$^{-2}$ and T$_{rot}$=29.9$\pm$0.6\,K. 
The C$_5$N$^-$/C$_7$N$^-$ abundance ratio is therefore $\sim$2.4, which is
similar to that found in TMC-1.

The detection of C$_7$N$^-$ in interstellar and circumstellar environments implies
that C$_7$N must also be present in both media if the anion is formed via electron
attachment to the radical. For the neutral species C$_7$N, ab initio calculations
by \citet{Botschwina1997} predict an electronic ground state $^2\Pi$ with a moderate
dipole moment of $\sim$1. However, due to the possible admixing of the ground
state with a low-lying $^2\Sigma$ state, which has a dipole moment of $\sim$3.6 \citep{Botschwina1997},
the real value of the dipole moment could be between those of both states. This mixing of
two low-lying electronic states was previously discussed for C$_5$N \citep{Cernicharo2008} and
C$_4$H \citep{Oyama2020}. Taking into account the observed intensities of C$_7$N$^-$ ($\mu$\,=\,7.545\,D) in TMC-1 and IRC\,+10216,
those of C$_7$N could be much lower due to the lower
dipole moment of the neutral and to the larger partition function of the radical
caused by the presence of two ladders, $^2\Pi_{1/2}$ and $^2\Pi_{3/2}$, each one exhibiting $\Lambda$-doubling
and hyperfine structure.
Adopting an averaged dipole  for C$_7$N of between those of the $^2\Pi$ and $^2\Sigma$ states of 2.3\,D
\citep{Botschwina1997} and a rotational partition function four times larger than for C$_7$N$^-$, the lines of the radical
could be $\sim$40\,$\times$\,N(C$_7$N$^-$)/N(C$_7$N) times weaker than those of C$_7$N$^-$.
Adopting a N(C$_7$N$^-$)/N(C$_7$N) similar to the N(C$_5$N$^-$)/N(C$_5$N), which 
is found to be $\sim$\,0.5 in IRC\,+10216 and TMC-1 
\citep{Cernicharo2008,Cernicharo2020}, the expected intensities will 
be approximately 20 times weaker for C$_7$N than for C$_7$N$^-$, which 
is well
below the present sensitivity of our data towards the two sources. 
We can also make a prediction of the expected intensities of C$_7$N based on those 
of the analogue radical C$_8$H, which has a similar partition function 
to C$_7$N (excluding the hyperfine structure), although a larger dipole 
moment of 6.5\,D. The lines of C$_7$N will have an intensity of 
N(C$_7$N)/N(C$_8$H)\,$\times$\,[($\mu$(C$_7$N)/$\mu$(C$_8$H)]$^2$ 
times those of C$_8$H; that is, $\sim$N(C$_7$N)/N(C$_8$H)/8. The lines of 
the $^2\Pi_{3/2}$ ladder of C$_8$H have an intensity of $\sim$5 mK in 
TMC-1 and of 4 mK in IRC\,+10216 at 31 GHz. Therefore, detecting C$_7$N in 
our data will require an abundance for this species of similar to or larger 
than that of C$_8$H, which
is very unlikely in view of the observed abundance ratios of 
other members of these two families of radicals. 
For example, the abundance ratios C$_3$N/C$_4$H and C$_5$N/C$_6$H are 
on the order of 0.1 in both IRC\,+10216 and TMC-1 
\citep{Cernicharo2000,Agundez2013}.

\begin{figure}
\centering
\includegraphics[width=0.95\columnwidth]{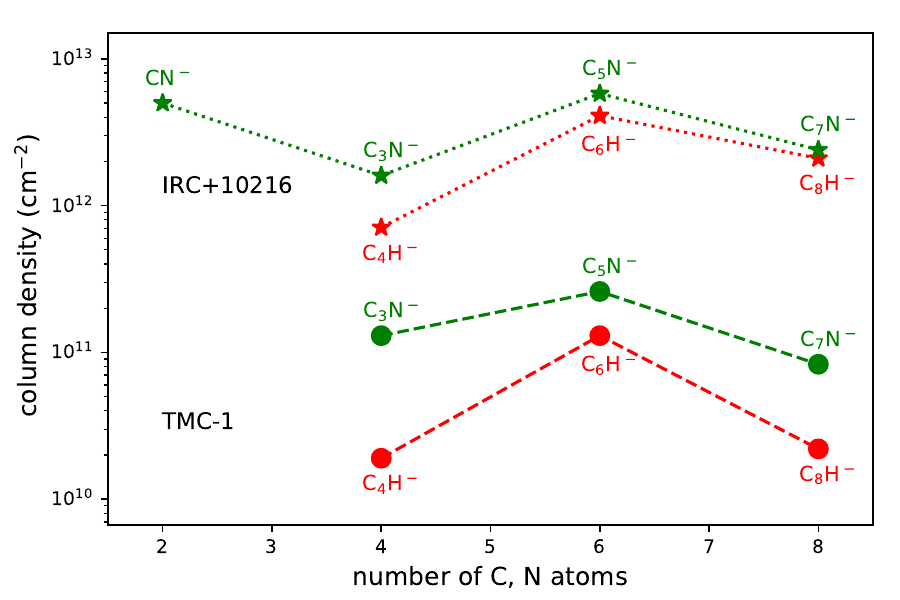}
\caption{Column densities of negative ions derived through observations toward TMC-1 and IRC\,+10216. Values are taken from this work, from Ag\'undez et al., in preparation, and from the literature \citep{Cernicharo2007,Cernicharo2020,Remijan2007,Thaddeus2008,Agundez2008,Agundez2010}.}
\label{fig:column_densities}
\end{figure}

The detection of C$_7$N$^-$ is in line with previous observational constraints 
of the series of hydrocarbon and nitrile anions C$_n$H$^-$ and C$_n$N$^-$ 
in which long carbon chains are favoured with respect to the short ones. 
Although we do not have access to the C$_7$N$^-$/C$_7$N abundance ratio, we 
expect this ratio to be in the range of 0.1-0.5, as found for other large 
anions, such as C$_8$H$^-$ and C$_5$N$^-$ 
\citep{Cernicharo2008,Cernicharo2020,Brunken2007,Remijan2007}. 
This scenario is in agreement with the proposed formation mechanism 
of negative ions from electron attachment to the corresponding radical 
\citep{Herbst1981}, which becomes increasingly efficient as the size of 
the chain increases, as calculated for the series of hydrocarbon radicals 
C$_n$H \citep{Herbst2008}. It would be interesting to compute the rate 
coefficient of electron attachment for the series of nitrile radicals 
C$_n$N. Calculated values for C$_3$N \citep{Petrie1997,Harada2008} 
and C$_5$N \citep{Walsh2009} indicate that, similarly to the case 
of hydrocarbons, the rate coefficient increases considerably as the 
size of the chain increases, although \cite{Khamesian2016} find much 
lower rate coefficients for C$_3$N and C$_5$N.

It is interesting to note that, although the anion-to-neutral ratio 
increases considerably with increasing chain length, the column densities 
of negative ions are on the same order along the series of hydrocarbons 
C$_n$H$^-$ and nitriles C$_n$N$^-$ (see Fig.~\ref{fig:column_densities}). 
This is interesting because if this behaviour is maintained for larger sizes, 
we could expect the anions C$_{10}$H$^-$ and C$_9$N$^-$ to be present in 
significant amounts, allowing  their detection in TMC-1 and IRC\,+10216 
at centimetre wavelengths. Moreover, looking at anions would allow us to probe the 
extent to which carbon chains can grow, and this would more straightforward than looking at 
neutrals. This is due to the fact that, for large carbon chains, the detection 
of the anion is favoured over the radical because of the collapse of the 
fine and hyperfine structure and eventually also because of the more favourable 
dipole moment, as perfectly illustrated in the case of C$_7$N$^-$ and 
C$_7$N.

\section{Conclusions}

We report in this work the detection of the anion C$_7$N$^-$ in TMC-1 and IRC\,+10216.
The lack of detection of the neutral C$_7$N is due to its much lower dipole moment and
larger partition function.

\begin{acknowledgements}
We thank ERC for funding
through grant ERC-2013-Syg-610256-NANOCOSMOS. M.A. thanks MICIU for grant 
RyC-2014-16277. We also thank Ministerio de Ciencia e Innovaci\'on of Spain (MICIU) for funding support through projects
PID2019-106110GB-I00, PID2019-107115GB-C21 / AEI / 10.13039/501100011033, 
and PID2019-106235GB-I00. 
\end{acknowledgements}

\onecolumn
\begin{appendix}
\section{Line parameters}
\label{line_parameters}
Line parameters for TMC-1 were obtained by fitting a Gaussian line
profile to the observed data. A window of $\pm$ 15 \kms\, around the v$_{LSR}$ of the source was
considered for each transition. The derived line parameters are given in Table \ref{line_parameters_tmc1}.
For IRC\,+10216, we used the SHELL method of the GILDAS$^3$ package which is well adapted to reproducing
the observed line profiles in circumstellar envelopes. A variable window of 50-100 MHz was used to
derive the line parameters in this source; these are given in Table \ref{line_parameters_irc}. 

\begin{table*}
\centering
\caption{Observed lines of C$_7$N$^-$ in IRC\,+10216.}
\label{line_parameters_irc}
\begin{tabular}{cccccccc}
\hline
Transition&$\nu_{obs}$$^a$    &$\nu_{obs}-\nu_{cal}$$^b$&$\int$ $T_A^*$ dv $^c$ & $T_A^*$(horn)\,$^d$ & $T_A^*(center)$\,$^e$ &$\sigma$$^f$ & Notes\\
\hline
27-26 & 31464.50$\pm$0.10& -0.11& 34.51$\pm$0.35 & 1.52& 1.01& &A \\
28-27 & 32630.00$\pm$0.10&  0.05& 43.94$\pm$0.44 & 1.66& 1.44& &B \\
29-28 & 33795.30$\pm$0.10&  0.02& 32.93$\pm$0.33 & 1.45& 0.97& & \\
30-29 & 34960.61$\pm$0.10&  0.00& 35.80$\pm$0.36 & 1.58& 1.06& & \\
31-30 & 36125.90$\pm$0.10& -0.04& 43.85$\pm$0.44 & 2.00& 1.35& & \\
32-31 & 37291.34$\pm$0.10&  0.07& 50.66$\pm$0.50 & 2.25& 1.50& &C \\
33-32 & 38456.60$\pm$0.10&  0.01& 50.64$\pm$0.51 & 2.24& 1.50& & \\
34-33 & 39621.88$\pm$0.10& -0.02& 49.44$\pm$0.49 & 2.00& 1.54& & \\
35-34 & 40787.30$\pm$0.10&  0.07& 47.66$\pm$0.48 & 2.11& 1.41& & \\
36-35 & 41952.50$\pm$0.10& -0.04& 49.49$\pm$0.49 & 2.18& 1.46& & \\
37-36 & 43117.90$\pm$0.10&  0.05& 54.01$\pm$0.54 & 2.39& 1.60& & \\
38-37 & 44283.10$\pm$0.10& -0.06& 49.51$\pm$0.50 & 2.01& 1.55& & \\
39-38 & 45448.50$\pm$0.10&  0.03& 42.65$\pm$0.43 & 1.89& 1.27& &D \\ 
40-39 & 46613.73$\pm$0.10& -0.04& 44.39$\pm$0.44 & 1.96& 1.31& &E \\ 
41-40 & 47779.10$\pm$0.10&  0.03& 49.21$\pm$0.49 & 2.17& 1.45& & \\
42-41 & 48944.30$\pm$0.10& -0.07& 44.66$\pm$0.45 & 1.82& 1.40& & \\
43-42 & 50109.70$\pm$0.10&  0.04& 43.83$\pm$13.0 & 1.78& 1.37& &F\\
\hline  
\hline                                                                                             
\end{tabular}
\tablefoot{\\
\tablefoottext{a}{Observed frequency assuming a v$_{LSR}$ of -26.5 \kms.}\\
\tablefoottext{b}{Observed minus calculated frequencies in MHz.}\\
\tablefoottext{c}{Integrated line intensity in mK\,km\,s$^{-1}$. Uncertainty is assumed to be
dominated by the calibration uncertainty of 10\%. This uncertainty has been multiplied
by a factor of 3 for the $J$=43-42 transition due to the poor atmospheric transmision
at 50.1 GHz.}\\
\tablefoottext{d}{Antenna temperature at the terminal velocity (horn) in milli Kelvin.}\\
\tablefoottext{e}{Antenna temperature at line center in milli Kelvin.}\\
\tablefoottext{f}{Root mean square noise of the data.}\\
\tablefoottext{A}{Blended with HCCCCC$^{13}$CCN. Line parameters can be still fitted.}\\
\tablefoottext{B}{Blended with C$_3$H. Line parameters can be still fitted.}\\
\tablefoottext{C}{Blended with HC$_7$N v$_{15}$.  Line parameters can be still fitted.}\\
\tablefoottext{D}{Blended with CH$_3$C$_3$N.  Line parameters can be still fitted.}\\
\tablefoottext{E}{Blended with a U feature.  Line parameters very uncertain.}\\
\tablefoottext{F}{Calibration very uncertain.}
}
\end{table*}
\begin{table*}
\centering
\caption{Observed lines of C$_7$N$^-$ in TMC-1.}
\label{line_parameters_tmc1}
\begin{tabular}{ccccccc}
\hline
Transition&$\nu_{obs}$$^a$    &$\nu_{obs}-\nu_{cal}$$^b$&$\int$ $T_A^*$ dv $^c$  &$\Delta$v\,$^d$& $T_A^*$\,$^e$ &$\sigma$$^f$\\

\hline  
27-26 & 31464.665$\pm$0.020& -5.0 &0.78$\pm$0.09&0.80$\pm$0.20& 0.97& 0.12$^g$\\ 
28-27 & 32630.005$\pm$0.010&  1.5 &0.65$\pm$0.08&0.76$\pm$0.05& 0.80& 0.10\\
29-28 & 33795.331$\pm$0.010& -3.3 &0.40$\pm$0.09&0.71$\pm$0.18& 0.52& 0.11\\
30-29 & 34960.667$\pm$0.010&  4.8 &0.49$\pm$0.05&0.90$\pm$0.12& 0.53& 0.08\\
31-30 & 36125.987$\pm$0.010& -0.3 &0.31$\pm$0.06&0.61$\pm$0.15& 0.48& 0.09\\
32-31 & 37291.308$\pm$0.020& -1.5 &0.24$\pm$0.05&0.53$\pm$0.11& 0.44& 0.09\\
33-32 & 38456.628$\pm$0.004&      &             &             &     &$\le$0.24$^h$\\
\hline                                                                                             
\end{tabular}
\tablefoot{\\
\tablefoottext{a}{Observed frequency assuming a v$_{LSR}$ of 5.83 \kms.}\\
\tablefoottext{b}{Observed minus calculated frequencies in kHz.}\\
\tablefoottext{c}{Integrated line intensity in mK\,km\,s$^{-1}$.}\\
\tablefoottext{d}{Line width at half intensity derived by fitting a Gaussian function to
the observed line profile (in km\,s$^{-1}$).}\\
\tablefoottext{e}{Antenna temperature in milliKelvin.}\\
\tablefoottext{f}{Root mean square noise of the data.}\\
\tablefoottext{g}{Line blended with a U feature at 31464.59 MHz.}\\
\tablefoottext{h}{3$\sigma$ upper limit. The frequency of this transition corresponds
to the predicted one from the constants given in Table \ref{rota}.}
}
\end{table*}

\end{appendix}
\end{document}